\documentclass[
$preprint,
reprint,
]{revtex4-2}

\usepackage{graphicx}
\usepackage{dcolumn}
\usepackage{bm}
\usepackage{xcolor}
\usepackage{placeins}
\usepackage{setspace}
\usepackage{bbold}

\newcommand{\etal}{\emph{et al.}}
\newcommand{\be}{\begin{equation}}
\newcommand{\ee}{\end{equation}}
\newcommand{\bfig}{\begin{figure}}
\newcommand{\efig}{\end{figure}}

\usepackage{lineno}
\usepackage{lipsum}

\begin{document}

\title{Anisotropic resistance with a 90-degree twist in a ferromagnetic Weyl semimetal, Co$_2$MnGa}

\author{Nicholas P. Quirk$^{1}$}
\author{Guangming Cheng$^{2}$}
\author{Kaustuv Manna$^{3}$}
\author{Claudia Felser$^{3}$}
\author{Nan Yao$^{2}$}
\author{N. P. Ong$^{1,\S}$}
\affiliation{
{$^1$Department of Physics, Princeton University, Princeton, NJ 08544, USA}\\
{$^2$Princeton Institute for the Science and Technology of Materials,}\\{Princeton University, Princeton, NJ 08544, USA}\\{$^3$Max Planck Institute for Chemical Physics of Solids, Nöthnitzer Str. 40, 01187 Dresden, Germany}}

\date{\today}
\begin{abstract}
Co$_2$MnGa is a ferromagnetic semimetal with Weyl nodal lines identified by ARPES~\cite{Belopolski}. We studied electrical transport in thin Co$_2$MnGa lamellae (10 $\times$ 10 $\times$ 0.4-5 $\mu$m) cut from single-crystals using a focused ion beam. These crystals exhibit an unexpected and highly unusual planar resistance anisotropy ($\sim$10$\times$) with principal axes that rotate by 90$^\circ$ between the upper and lower faces. Using symmetry arguments and simulations, we find that the observed resistance anisotropy resembles that of an isotropic conductor with anisotropic surface states that are impeded from hybridization with bulk states. The origin of these states awaits further experiments that can correlate the surface bands with the observed 90$^\circ$-twist geometry.
\end{abstract}

\maketitle
In a Weyl semimetal (WSM), the breaking of time-reversal invariance or inversion symmetry leads to the splitting of each Dirac node into two Weyl nodes that are separated in $\bf k$ space and have opposite chiralities. The surface projections of these Weyl nodes act as terminations for topologically protected Fermi arcs~\cite{Armitage}. Transport experiments on single crystals have observed the chiral anomaly involving bulk states in parallel electric and magnetic fields~\cite{Xiong, Huang, Zhang, Hirschberger} and Fermi-arc-mediated transport has been inferred through quantum oscillations in thin crystals~\cite{Moll,
ZhangC1, Zheng, ZhangC2, Schumann}. However, so far novel transport phenomena in WSMs have only been observed when topological states are brought into the quantum limit at cryogenic temperatures in a strong magnetic field.

\begin{figure*}
    \centering
 \includegraphics[width=\textwidth]{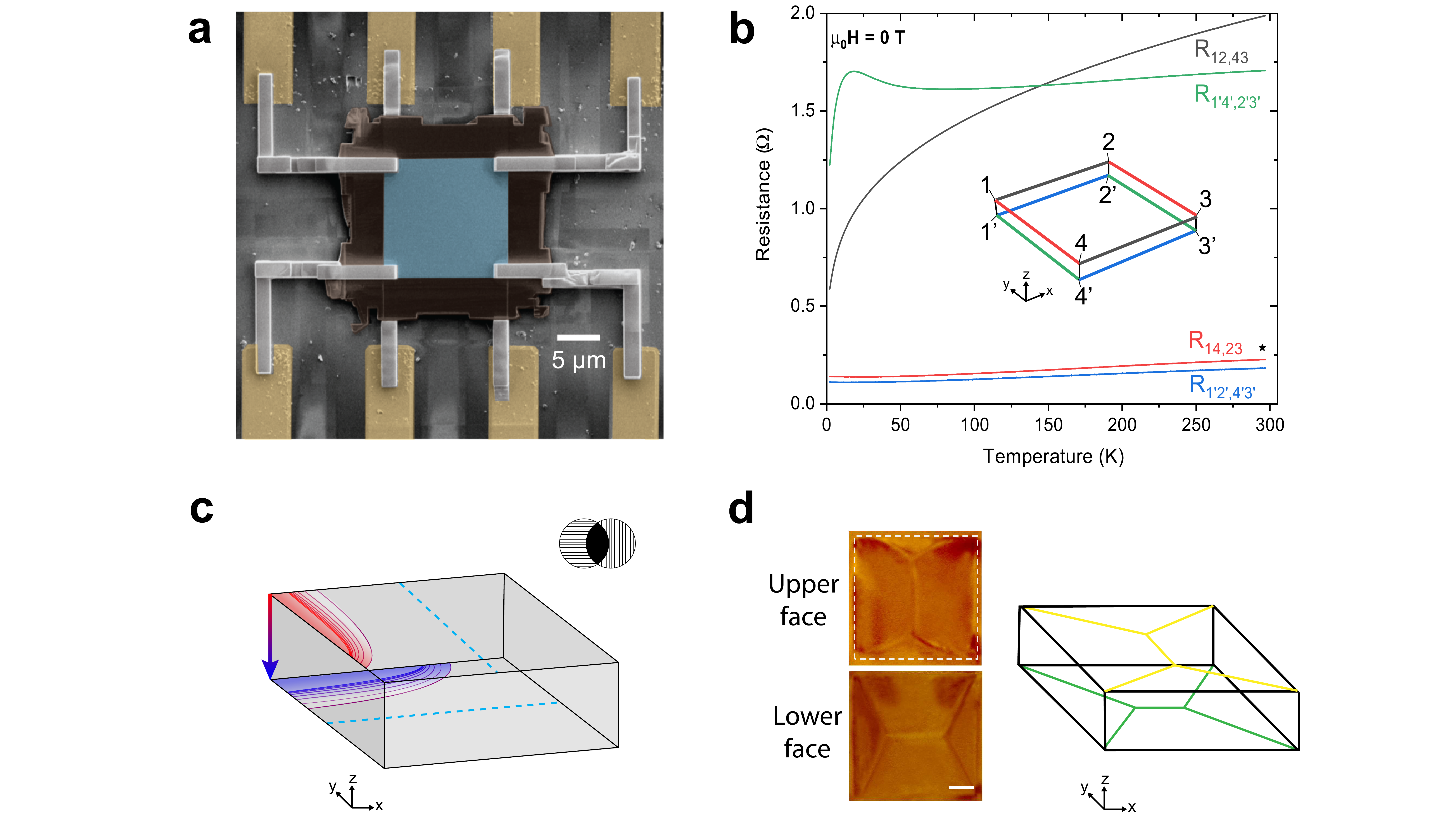}
    \caption{ {\bf Co$_2$MnGa thin-plate devices. a.} Top-view false-color SEM micrograph of Sample N2 (12 $\times$ 12 $\times$ 0.4 $\mu$m). The central blue region is the Co$_2$MnGa lamella. Insulating, amorphous carbon (dark brown) is added with FIB to protect the sides of the device and prevent electrical shorting of the platinum contacts. On the underside (not visible), small Pt nibs are added to facilitate electrical connections from the side. The sample is then mounted onto an SiO$_2$ substrate and more Pt (grey) is added to connect the corners to pre-patterned gold electrodes (yellow).  {\bf b.} The temperature dependences of the four-point resistances on the upper/lower faces ($H$ = 0 T). The high-conductance-axes (upper =  $\bf{\hat{y}}$, lower = $\bf{\hat{x}}$) resistances have metallic profiles that match bulk Co$_2$MnGa. (The isotropic equivalent resistance is denoted with a black star at $T$ = 290 K.) The low-conductance-axes (upper =  $\bf{\hat{x}}$, lower = $\bf{\hat{y}}$) resistances have anomalous temperature dependences that decay slowly at high temperatures but sharply below $T$ = 10 K. The inset diagram depicts the corner-contact labeling scheme used throughout to identify each $R_{ij,kl}$. {\bf c.} A cartoon depicting the 90$^\circ$-twist planar anisotropy in a 3D lamella. The dashed blue lines indicate the high-conductance axes. The anisotropy is such that when current is directed along the thin vertical edges, a surface potential arises that is strongly distorted in opposite lateral directions on the upper/lower faces. This effect is reminiscent of how crossed polarizers block light (upper-right inset). {\bf d.} Twisted Landau flux-closure patterns in thin lamellae. On the left are magnetic force micrographs of the flux-closure domain-wall patterns that arise in a sample with thickness $d$ = 1300 nm oriented to $\bf{\hat{x}}$ $\parallel\langle 110 \rangle$, $\bf{\hat{y}}$ $\parallel\langle \overline{1}10 \rangle$, the same as Sample O1. The upper (lower) image is the pattern that arises on the top (bottom) face. The diagram on the right depicts how the patterns appear on a 3D lamella. Scale bar = 1 $\mu$m. }
    \label{fig:fig1}
\end{figure*}

Here we report the observation of a most unusual electronic feature in 
thin (0.4--5 $\mu$m) plate-like crystals of the ferromagnetic Weyl nodal-line semimetal Co$_2$MnGa. All 6 crystals investigated  display a large conductance anisotropy on both of the broad faces (the $a$-$b$ plane). However, the principal axes are invariably rotated by 90$^\circ$ between the upper and lower surfaces (Fig. \ref{fig:fig1}). Using symmetry arguments and simulations, we infer that the anisotropies originate from surface states $|{\bf q}\rangle$ that are protected from hybridization with bulk states $|{\bf k}\rangle$, i.e., surface-to-bulk charge transfer is mediated by a transfer matrix $t({\bf q,k})$ that is strongly anisotropic. The intrinsic 90$^\circ$ twist impedes charge flow in a way reminiscent of how crossed polarizers block light transmission (but the physics is different). The obstruction leads to transport anomalies that satisfy an explicit roto-inversion symmetry $C_4I$, and are observable up to macroscopic lengths (5 $\mu$m) at room temperature.

Co$_2$MnGa forms a face-centered cubic lattice, space group $Fm\overline{3}m$ (no. 225) and is a soft ferromagnet ($H_C$ $\simeq$ 30 Oe) with Curie temperature $T_C$ = 690 K. Two majority-spin bands near the Fermi energy form Weyl lines that are twofold degenerate due to mirror symmetry \cite{Belopolski}. Additionally, it exhibits giant anomalous Hall and Nernst responses that have been related to the large Berry curvature at the Weyl nodal lines~\cite{Belopolski, Guin, Sakai, Reichlova, ZhangY}. Dispersionless, 2D drumhead surface states have also been observed in Co$_2$MnGa by photoemission \cite{Belopolski}. 

The present transport experiments were motivated by images of anomalous remanent magnetic domain patterns obtained by magnetic force (MFM) and Lorentz transmission electron microscopy (LTEM)~\cite{ChengYao}. Using a focused ion beam (FIB) microscope, we cut thin, square-shaped lamellae of thickness $d$ $\lesssim$ 1 $\mu$m from as-grown Co$_2$MnGa single crystals. Each sample contains just a few domains, which form the familiar Landau flux-closure pattern~\cite{Landau} on each face (Fig. \ref{fig:fig1}d). However, the pattern on the lower face is rotated by 90$^\circ$ with respect to that of the top. This 90$^\circ$ twist remains in lamellae with $d$ down to 100 nm. 

In a ferromagnetic WSM, the bulk magnetization $\bf M$ couples to the Weyl nodes with opposite signs. Inhomogeneities in $\bf M$ such as domain walls are predicted to host localized charge and chiral equilibrium currents ~\cite{Grushin, Araki1, Araki2}. Initially, we tried to identify the anisotropies in the resistance matrix $R_{ij,kl}$ (defined below) with conductances along domain walls. However, Co$_2$MnGa is a soft ferromagnet with a coercive field of just 30 Oe. We find that magnetotransport measurements on these thin crystals ($d$ = 0.4--5 $\mu$m) show clear evidence for the erasure of domain walls at low field. As shown below, the Hall resistances feature sharp hysteretic peaks at the coercive field and saturate to a constant slope at 1 T.  Furthermore, each $R_{ij,kl}$ has the same overall weak magnetoresistance ($<5\%$ at 3 T).  Because all elements in $R_{ij,kl}$ are nearly field-independent, we eventually concluded that the observed anisotropies are intrinsic to the twisted electronic band structure rather than arising from domain walls. Although the 90$^\circ$-twisted flux-closure patterns in these thin lamellar samples are of interest, a detailed study of the remanent domains will be reported elsewhere~\cite{ChengYao}.

We investigate the electrical anisotropies by measuring the non-local resistance matrix $R_{ij,kl}$ = $V_{kl}/I_{ij}$ where $V_{kl}$ is the voltage measured between contacts $(k,l)$ with current applied to contacts $i$ and $j$. We label the four vertices of the upper (lower) face as 1$\cdots$4 (1'$\cdots$4'), as shown in the inset of Fig. \ref{fig:fig1}b. All resistance measurements are true four-probe ($i,j$ are distinct from $k,l$). To distiguish the anomalous, intrinsic anisotropies from trivial geometrical effects, we compare the measured $R_{ij,kl}$ against the values calculated for a metallic slab (of identical shape and size) with \emph{isotropic} resistivity (130 $\mu\Omega$-cm at 290 K~\cite{Belopolski}). The calculation was performed using the software package COMSOL. We refer to this as the ``isotropic equivalent." Our present analysis is focused on measurements of one device (Sample O1) with dimensions 10 $\times$ 10 $\times$ 1 $\mu$m and edges oriented to the crystal lattice as $\mathbf{\hat{x}}\parallel\langle110\rangle$, $\mathbf{\hat{y}}\parallel\langle\overline{1}10\rangle$, and $\mathbf{\hat{z}}\parallel\langle001\rangle$.  Similar results in 5 other devices (all exhibiting the 90$^\circ$ twist) are described in the Supplementary Materials. 

Figure 3a depicts the resistance anisotropy that arises when current is directed in the $x$-$y$ plane. All measurements are at $T$ = 290 K and $H$ = 0 T. With the pair of directed current contacts $ij$ (written as the vector $\vec{ij}$) $\parallel \mathbf{\hat{y}}$ on the upper face, the resulting voltage in that plane agrees with the expected value for the isotropic equivalent: $R_{14,23}$ = 0.2 $\Omega$ (experimental) $\simeq$ 0.29 $\Omega$ (isotropic simulation). However, with $\vec{ij}$ $\parallel \mathbf{\hat{x}}$, the resistance is $\sim 10\times$ greater: $R_{12,43}$ = 1.9 $\Omega$. Yet, in both cases the voltages on the lower face remain largely undistorted from their expected isotropic values (blue arrows in Fig. \ref{fig:fig1}a). If we direct the current in the lower face, a planar anisotropy arises that is rotated by 90$^\circ$. With $\vec{ij}$ $\parallel \mathbf{\hat{x}}$, $R_{1'2',4'3'}$ = 0.18 $\Omega$ agrees with the isotropic equivalent, but $R_{1'4',3'2'}$ =  2.5 $\Omega$ ($\parallel \mathbf{\hat{y}}$) is an order of magnitude larger. Thus, the crystal has a unique resistance anisotropy with a high conductance axis $\parallel$ $\bf{\hat{y}}$ ($\bf{\hat{x}}$) on the upper (lower) face. When current is directed along these axes, all of the $R_{ij,kl}$ agree with the values predicted by the isotropic equivalent. Only when the current is directed in a low-conductance axis is an anomalous resistance observed, and then only across the parallel edge on that face.

Out-of-plane currents ($\vec{ij}\parallel{\bf \hat{z}}$) bring out the most dramatic features of the anisotropy (Fig. \ref{fig:fig3}b). In the isotropic equivalent simulation, the 7 distinct $R_{ij,kl}$ for $I$ applied along each vertical edge are $\simeq$ 10$^{-7}$ $\Omega$ (consistent with the non-local resistance in a high-conductivity metal). However, in the experiment four of these resistances are $\sim$6 orders of magnitude greater, about 1 $\Omega$. For example, with $ij$ = $11'$ (top-left panel in Fig. \ref{fig:fig3}b) the voltages on the adjacent vertical edges are 1.3 $\Omega$ ($kl$ = $22'$) and 1.0 $\Omega$ ($kl$ = $44'$). Along the diametrically opposed edge ($kl$ = $33'$), however, the resistance is very small, about -1 m$\Omega$. Additionally, the in-plane voltages alternate between $\simeq$ 1 $\Omega$ and $\simeq$ 10 m$\Omega$---on edges of the crystal that are separated by just 1 $\mu$m. 

Together these measurements map out a surface potential $V({\bf r}_s)$ with a robust 90$^\circ$-twist symmetry. The diagrams in each of the 4 panels of Figure \ref{fig:fig3}b (corresponding to $\vec{ij}$ along each vertical edge) can be mapped into each other by a rotation about the $z$ axis by an angle $\phi = n\pi/2$ ($n\in\mathbb{Z}$) followed by an inversion about the origin (geometric center of the crystal). Thus, we identify the symmetry group of the observed 90$^\circ$ twist as $C_{4z}I$. This procedure can also be applied to the in-plane resistance configurations shown in Figure \ref{fig:fig3}a.

The 90$^\circ$-twist anisotropy is largest at room temperature but still present at $T$ = 2 K. Additionally, there is a marked difference in the temperature dependences of the low- and high-conductance-axis in-plane resistances (Fig. \ref{fig:fig1}b). The high-conductance axis resistances ($R_{14,23}$ and $R_{1'2',4'3'}$) have metallic profiles matching bulk Co$_2$MnGa. The low-conductance-axis resistances, however, have anomalous temperature dependences. $R_{12,43}$ decays gradually at high temperature but sharply below 10 K and $R_{1'4',2'3'}$ has a non-monotonic dependence with a peak at the same characteristic temperature. We have observed similar temperature dependences in every sample (see Supplementary Materials).

As mentioned, the anisotropies are virtually unaffected by applied magnetic fields. With $\bf H$ $\parallel$ $\bf \hat{z}$, we measured the magnetoresistance (MR) and Hall effect using the contacts on both faces of the crystal. Figures \ref{fig:fig3}a and \ref{fig:fig3}b show the MR in Sample N2 ($d$ = 0.4 $\mu$m) at $T$ = 20 K and Fig. S3a shows the MR in Sample O1 ($d$ = 1 $\mu$m) at $T$ = 50 K. The two plots are qualitatively identical. Above $H$ = 1.3 T, each MR decreases monotonically, reaching $\simeq$ -10 m$\Omega$ at $H$ = 3 T (Sample N2, Fig. \ref{fig:fig3}a). The applied $H$ has a uniform effect on each resistance, regardless of whether the current is directed in a high- or low-conductance axis: the intrinsic anisotropy remains virtually unchanged. Below $H$ = 1.3 T, we observe slight ($x$-$y$) anisotropic MR that is independent from the underlying 90$^\circ$ twist. Similar weak anisotropic MR has been observed in Co$_2$MnGa thin films~\cite{Tong} where it has been attributed to domain-dependent scattering. 

The field-antisymmetric MR is more puzzling. With $\vec{ij}\parallel{\bf \hat{y}}$ (but not $\parallel{\bf \hat{x}}$), the MR has a large $H$-antisymmetric component that shares the same knee at $H\simeq$ 1 T as the anomalous Hall resistance (Fig. \ref{fig:fig3}c). Exchanging the current and voltage contacts changes the sign of the antisymmetric slope, $R_{ij,kl}({\bf H}) = R_{kl,ij}(-{\bf H})$, in agreement with the reciprocity theorem~\cite{SBSS}. The origin of this antisymmetric MR is unknown. Additionally, the antisymmetric MR presents an intriguing pattern when ${\bf H}$ is directed in the $x$-$y$ plane (Fig. S4 in Supplementary Materials).

Hall-effect measurements show sharp, hysteretic peaks near the coercive field ($\simeq$ 30 Oe), providing additional evidence for the low-field removal of domain walls (Fig. \ref{fig:fig3}c). Also, although the raw transverse resistances, e.g., $R_{13,42}$ and $R_{4'2',3'1'}$, reflect the intrinsic 90$^\circ$-twist anisotropy at zero $H$, the true Hall signal (Van der Pauw requires averaging the two transverse configurations) is identical on either face (Fig. \ref{fig:fig3}d). Furthermore, after scaling by the lamella thickness, the anomalous Hall resistivity $\rho_{yx}$($H$ = 3 T) agrees with that of the as-grown crystals (6 $\mu\Omega$-cm at 2 K)~\cite{Belopolski}.

The 90$^\circ$-twist anisotropy arises in every thin-plate crystal we have studied, including samples as thick (thin) as 5 $\mu$m (0.4 $\mu$m), with a 2:1 aspect ratio, and with edges oriented to the ($\langle100\rangle$, $\langle 010\rangle$) lattice directions  (Table 1 in Supplementary Materials). In each sample, the high-conductance-axis resistances agree with the expected values for an equivalent isotropic slab whereas the low-conductance-axis resistances are fixed at the few-$\Omega$ level.\\

\noindent\emph{Symmetry constraints}\\\noindent
Once we discovered empirically that $R_{ij,kl}$ satisfies $C_4I$ symmetry, the configurations needed to characterize the anisotropy were reduced considerably. With current applied out-of-plane ($\vec{ij}\parallel\bf\hat{z}$) there are 7 distinct configurations, whereas for planar current $\vec{ij}\parallel\bf\hat{x}$, $\bf\hat{y}$ there are 14 (within the symmetry group). In the planar configurations we further exclude the four transverse voltage pairs which may be inferred by applying the closed-loop theorem. Hence 13 measurements suffice to characterize $R_{ij,kl}$.

Symmetry leads to a more interesting constraint on the out-of-plane transport. Arguably, the most singular feature is the appearance of near-zero resistance on the vertical edge diametrically opposite to the current contacts, which is juxtaposed between extremely large resistances across the two remaining vertical edges. For e.g. with current $I_{11'}$, $V_{33'}\simeq 0$ whereas $R_{11',22'},\;R_{11',44'}\simeq 1\; \Omega$. The vanishing $V_{33'}\simeq 0$ is a consequence of the $C_4I$ symmetry independent of a microscopic model. 

We consider the electric potential function $V({\bf r}_s)$ on the two side surfaces 233'2' and 344'3' with the current injected at 1 and drained at 1' (${\bf r}_s$ locates a point on either side surface). For convenience, we map $\bf r$ onto the 2D plane with coordinates $(u,z)$ and origin at the mid point of 33'. The $C_4I$ symmetry implies that, on the $(u,z)$ plane, the potential distribution must satisfy the two constraints (see Supplementary Materials)

\be
V(u,z) =  -V(-u,-z), \quad V(u,z) \ne V(-u,z).
\label{Vuz}
\ee
The first is a consequence of $C_{2d}$ (rotation by $\pi$ of the $(u,z)$ plane about the normal through its origin). The second is a consequence of current flow in a potential field. With the assumed current, $V$ on the upper edge 32 must be higher than that on the lower edge 3'2' (see Supplementary Materials). The simplest function satisfying Eq. \ref{Vuz} has the form
\be
V(u,z) = u(z.{\rm sign}(u)-d/2).
\ee
Hence symmetry constrains $V=0$ all along the contact pair 33' ($u=0$).\\

\noindent\emph{3D Simulation}\\\noindent
We have obtained considerable insight into the observed $R_{ij,kl}$ with finite-element electrostatic simulations using COMSOL. In loose analogy with crossed polarizers in optics, we may regard the 90$^\circ$ twist as a mechanism that obstructs charge transport along $\bf\hat{z}$ with observable effects over macroscopic length scales (sample thickness $d$ up to 5 $\mu$m). This seems physically possible only if the surface states $|{\bf q}\rangle$ in Co$_2$MnGa are protected against hybridization with the bulk states $|{\bf k}\rangle$, and charge transfer between the surface and the bulk proceeds by a hopping matrix $t({\bf q,k})$. On the upper face, we express the anisotropic conductivity as
\begin{equation}
\sigma^s_{yy} = \sigma_0, \quad \sigma^s_{xx} = \sigma_0/\alpha, \quad (\alpha>1).
\label{syy}
\end{equation}
The transfer matrix $t({\bf q,k})$ should reflect this anisotropy as well. (For e.g., if the anisotropy is expressed as an anisotropic surface band mass, $m_y\sim m_0$ and $m_x = \alpha m_0$ where $m_0$ is the bulk mass, Fermi wavevector mismatch strongly suppresses the transfer rate for $\bf q\parallel\hat{x}$.) 

Heuristically, we may simulate the effect of an anisotropic $t({\bf q,k})$ by replacing the strictly 2D surface states by an ultrathin layer of a 3D anisotropic conductor that has a conductivity tensor $\sigma^s_{ij} = \mathrm{diag}[\sigma_0/\alpha,\sigma_0,\sigma_0/\beta]$ with $\beta\gg\alpha>1$. When $\beta$ is very large ($10^4 - 10^5$), the vanishing $z$-axis conductivity simulates an overall weak amplitude for the transfer matrix.

Using the Electric Currents COMSOL package, we represent the crystal as a lamellar slab of size 10 $\times$ 10 $\times$ (1 - 2$\delta)$ $\mu$m and isotropic conductivity $\sigma_0$, sandwiched between two ultrathin sheets of conductors with conductivity $\sigma^s_{ij}$ and thickness $\delta$. For $\sigma_{0}$ we used the value $7.7\times 10^5$ ($\Omega$m)$^{-1}$, measured in bulk crystals at 290 K~\cite{Belopolski}. We enforce continuity of the electric field across the interfaces and add point-contact probes to each of the 8 vertices. With specific choices of $I_{ij}$ and $V_{kl}$, we then simulate $R_{ij,kl}$ by calculating the voltage drops $V(\bf{r})$ across the edges of the composite slab. We choose $\alpha$ and fine-tune $\beta$ to match the experimental low-conductance-axis resistance $R_{12,43}$ = 1.9 $\Omega$ within 0.5\%. We then evaluate how well the remaining $R_{ij,kl}$ outputs agree with the experimental values. Due to the $C_4I$ symmetry it suffices to simulate only $\frac14$ of the possible four-point resistance configurations. Although convergence problems preclude simulations with $\delta$ $<$ 10 nm, we find that this picture can recreate the observations with surprising accuracy (the best agreement is achieved in the limit $\delta$ = 10 nm).

This simple model captures the key aspects of the observed 90$^\circ$-twist anisotropy in thin crystals. With $\delta$ = 10 nm, we run the simulation with $\alpha$ = $\{$10, 100, 500, 1000$\}$ and record the resulting $R_{ij,kl}$ (Figs. \ref{fig:fig4}a and \ref{fig:fig4}b). The $\beta$ required to match the experimental low-conductance-axis resistance ($R_{14,23}$) for each $\alpha$ are $\{$5.76, 4.31, 3.65, 3.4$\}$ $\times$ 10$^4$, respectively. With these ($\alpha$, $\beta$), the model succeeds in holding the high-conductance-axis resistances to their experimental values, which match the isotropic equivalent, while also generating anomalous resistances that agree with the observed 90$^\circ$-twist. It nearly captures the large 1-$\Omega$ voltages for current directed along the thin vertical edges, reaching $\simeq$ 0.8 $\Omega$ for $\alpha$ = 1000. Furthermore, it preserves the vanishingly small resistances that arise in this configuration, which are enforced by the $C_4I$ symmetry. It fails to capture the full magnitude of the voltage directly beneath the applied-current edge, undershooting $R_{14,1'4'}$ and $R_{12,1'2'}$ by $\sim$50\%, however. As $\alpha$ is increased the agreement improves, but stills falls short ($\alpha$ = 1000 is close to the asymptotic value). This issue could perhaps be rectified by a more elaborate model where the role of the narrow side faces is taken into account.

The simulation results are only weakly dependent on $\alpha$, but critically sensitive to the out-of-plane anisotropy $\beta$. For example, the simulated $R_{ij,kl}$ that arise for $\alpha$ = 10 are within 10\% of those for $\alpha$ = 1000. However, we find that no anisotropy arises for $\beta$ $<$ 10$^4$ (Fig. \ref{fig:fig4}c). Below this critical value, the surface low-conductance-axis resistances are short-circuited by the isotropic bulk. Thus, the surface conductances must be partially isolated from the bulk in order to generate the observed 3D twisted anisotropy. This is most evident when the current is applied vertically ($\vec{ij}\parallel{\bf\hat{z}}$). The 1-$\Omega$ resistances on neighboring vertical edges are a consequence of the strong planar distortion of the current in the ultrathin surfaces. Figure \ref{fig:fig4}d shows the highly distorted $V(\bf{r})$ contours. Due to the large $\beta$, the current density cannot easily flow directly to the drain at $1'$. A portion spreads out laterally in the surfaces before finding the path of least resistance through the bulk. The large magnitudes of the voltages at $22'$ and $44'$ are thus a measure of the weak amplitude of the transfer matrix $t({\bf q,k})$ for surface states to enter the bulk.

As the thickness of the anisotropic surface sections is increased, the simulated $R_{ij,kl}$ diverge from the experimental values. Figures \ref{fig:fig5}a and \ref{fig:fig5}b show the results of a series of simulations with $\alpha$ = 100 for $\delta$ = 10, 100, 300, and 500 nm (this thickest model consists of just two anisotropic slabs). With larger $\delta$, the simulated $R_{ij,kl}$ with $\vec{ij}$ directed along a high-conductance-axis no longer match the isotropic equivalent as they do in the experiment. Similarly, the voltage drops in the lower face, e.g., $R_{14,2'3'}$ and $R_{12,4'3'}$, also increase from the expected isotropic equivalent values. Finally, with $\vec{ij}\parallel{\bf\hat{z}}$, the anomalous resistances (e.g., $R_{11',22'}$) diminish rapidly and the vanishing-resistance edges increase (e.g., $R_{11',23}$). The thick-surface models do achieve better agreement for the resistances of the side faces, e.g., $R_{14,1'4'}$ and  $R_{12,1'2'}$, but because of their failures in every other regard, appear to describe a different system than what is observed in the experiment.\\

\noindent\emph{Implications and outlook}\\\noindent
Initially, our working assumption was that the anisotropies observed in $R_{ij,kl}$ arise from current paths confined to domain walls. However, as discussed above, we abandoned this viewpoint following the Hall and magnetoresistance measurements which show that all $R_{ij,kl}$ values remain nearly unchanged (at the few $\%$ level) in $H$ as large as 9 T. Our conclusion is that the anisotropies in $R_{ij,kl}$ are intrinsic to the electronic band structure when subject to the 90$^\circ$ twist. The emergence of this twist in all lamellar crystals investigated shows that it reflects an intrinsic instability in Co$_2$MnGa. In effect the bulk electronic bands spontaneously undergo the 90$^\circ$ twist in thin crystals.

These unusual effects raise a number of open questions. What is the microscopic mechanism that underlies this rare instability? Is the twist driven by the topologically non-trivial nature of the bulk states? Where is the energy gain that offsets the cost incurred by the twist? What gives rise to the large conductance anisotropy of the protected surface states, and why does the Landau magnetic domain pattern lock to the conductance anisotropy axes? 

Quite apart from these questions, we have in these crystals an opportunity to detect the presence of anisotropic surface states that are protected from hybridizing with high-conductance isotropic bulk states. We observed that the 90$^\circ$-twist impedes the out-of-plane charge current in a complex way consistent with the underlying $C_4I$ symmetry. For current applied to say the vertices $11'$, the voltage drops across both adjacent vertices $22'$ and $44'$ are extremely large ($R_{11',22'}\sim 1 \;\Omega$) whereas that across the opposite vertices $33'$ is extremely small ($R_{11',33'}\sim 1\;$ m$\Omega$). Although these anisotropies are reproducible in our simulation, the underlying physics for the appearance of a 1-$\Omega$ non-local resistance in a metallic crystal of sub-micron thickness requires further investigation.

\newpage
\vspace{1cm}
{\bf Acknowledgement}
We have benefited from discussions with L. Balents, B.A. Bernevig, J. Herzog-Arbeitman, and B. Lian. The research was funded by 
the U.S. Department of Energy (DE-SC0017863) and the Gordon and Betty Moore Foundation's EPiQS initiative via grant GBMF9466 (to N.P.O.).

\vspace{3mm}
\noindent
{\bf Author contributions}\\
N.P.Q., G.C., N.Y. and N.P.O. conceptualized and designed the experiment. Device fabrication was carried out by G.C. and N.P.Q. N.P.Q. performed all electrical measurements as well as the finite-element simulations. The crystals were grown by K.M. and C.F. Analyses of the data were carried out by N.P.Q. and N.P.O. who jointly wrote the manuscript with input from all authors.

\vspace{3mm}
\noindent
{\bf Competing financial interests}\\
The authors declare no competing financial interests.

\vspace{3mm}
\noindent
{\bf Additional Information}\\
Supplementary Materials is available in the online version of the paper.

\vspace{3mm}
\noindent
{\bf Correspondence and requests for materials}
should be addressed to N.P.Q. or N.P.O.

\nocite{*}
\bibliography{}

\begin{figure*}
    \centering
    \includegraphics[width=0.8\textwidth]{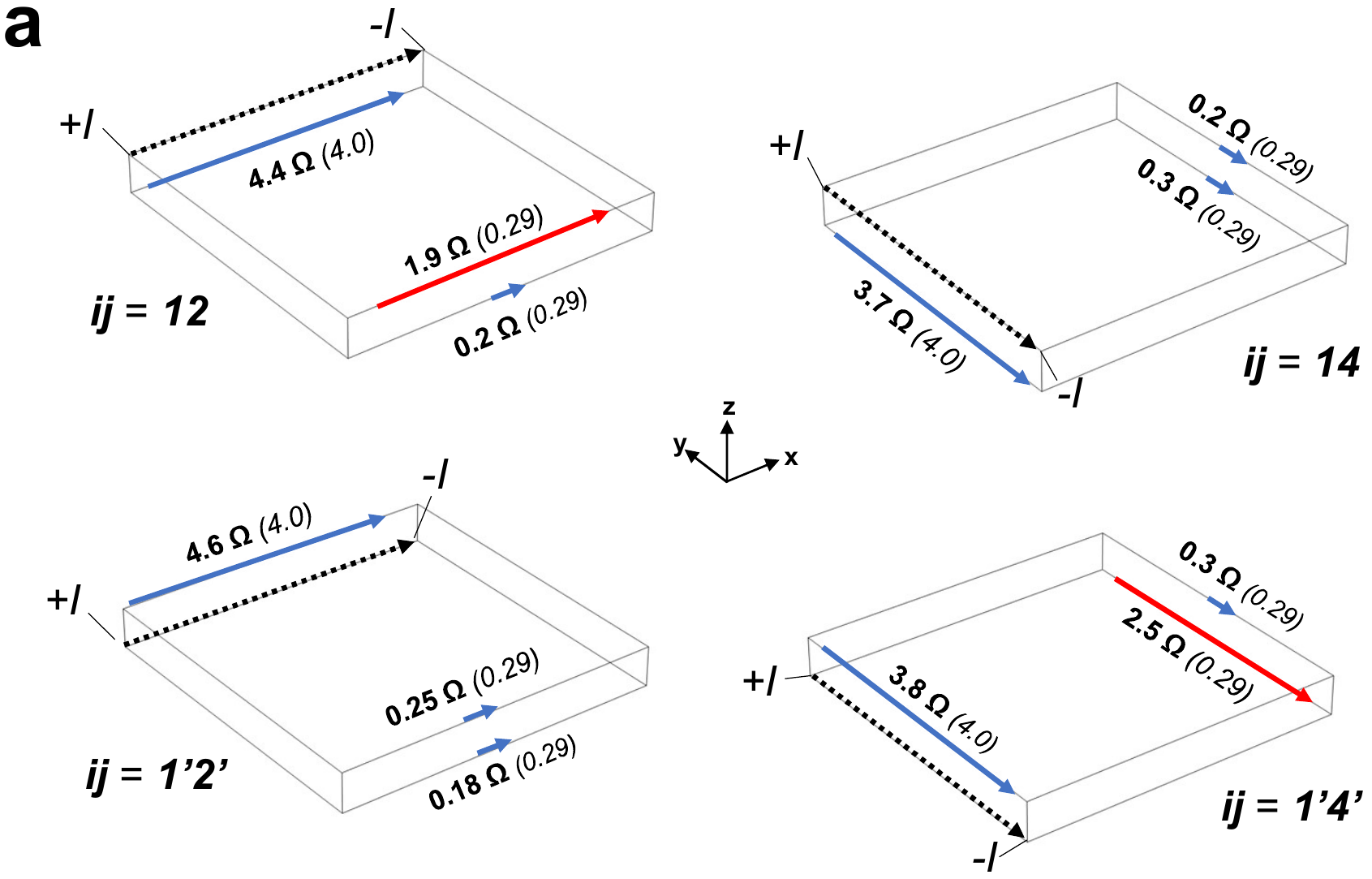}
   \includegraphics[width=0.8\textwidth]{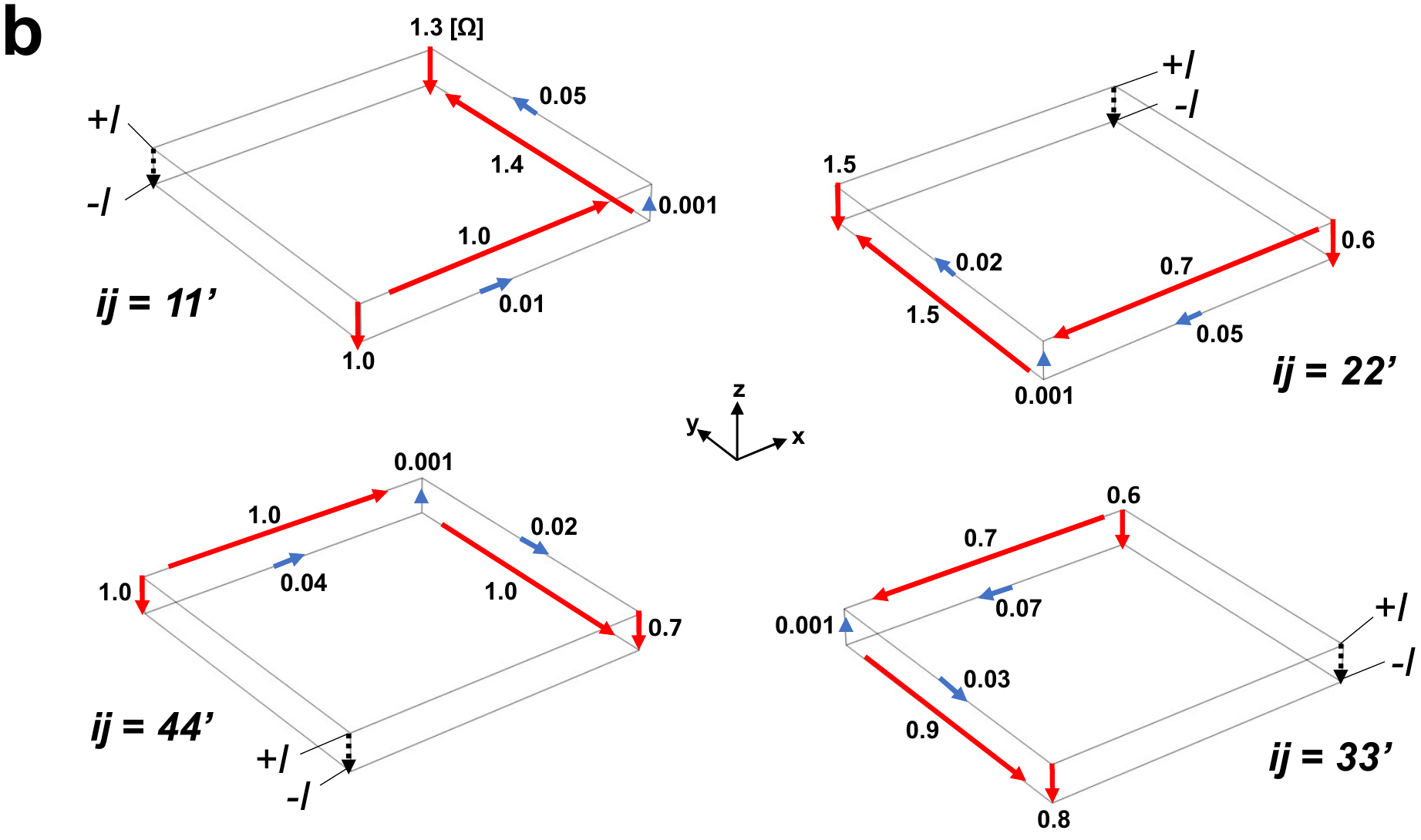}
    \caption{{\bf 90$^\circ$-twist resistance anisotropy. a.} Four-point resistances measured in Sample O1 at 290 K ($H$ = 0 T) with the current directed in the $x$-$y$ ($a$-$b$) plane. Each panel is identified by the source and drain current contacts ($ij$) according to the diagram in the inset of Fig. \ref{fig:fig1}a. For each $I_{ij}$, the voltage drop along every other parallel edge, $V_{kl}$, is measured. The value of $R_{ij,kl}$ = $V_{kl}$/$I_{ij}$ is recorded in $\Omega$. The length and color of the arrows represent the magnitude of the resistance and the agreement with the isotropic equivalent, respectively; the blue arrows align with the isotropic values and the red arrows are anomalously large. The isotropic equivalent expected values are given in parenthesis. When the current is directed along a high-conductance axis (upper = $\bf{\hat{y}}$, lower = $\bf{\hat{x}}$), all of the resulting $R_{ij,kl}$ agree with the isotropic equivalent (blue arrows). {\bf b.} When current is directed along the 1 $\mu$m vertical edges a highly distorted potential landscape arises. The expected values in the isotropic equivalent are all $\simeq$ 10$^{-7}$ $\Omega$. Thus, the red arrows correspond to resistances that are anomalous by $\sim$6 orders of magnitude. The blue arrows correspond to vanishingly small voltage drops, which we show in the \textit{Symmetry constraints} section to be enforced by the robust $C_4I$ symmetry of the 90$^\circ$ twist. }
    \label{fig:fig2}
\end{figure*}

\begin{figure*}
    \centering
 \includegraphics[width=\textwidth]{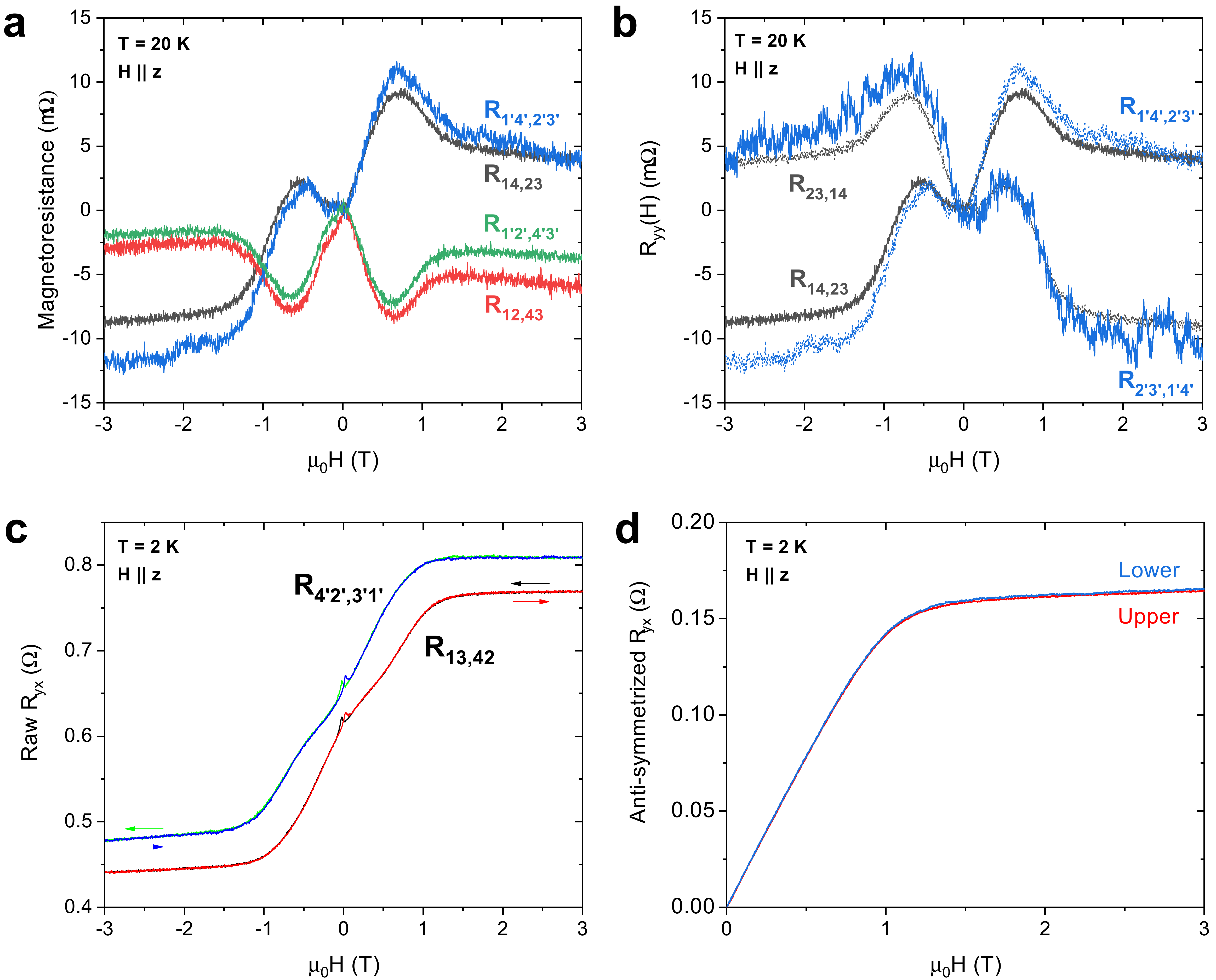}
    \caption{{\bf Magnetoresistance and Hall effect ($\bf H \parallel \bf \hat{z}$)}. {\bf a.} The magnetoresistance ($R(H)$ -- $R(H = 0)$) of the four in-plane resistances at $T$ = 20 K in Sample N2 (400 nm thick). The zero-$H$ values that have been subtracted are much larger than $\Delta R(H)$: $R_{14,23}$ = 0.31 $\Omega$, $R_{12,43}$ = 1.71 $\Omega$, $R_{1'4',2'3'}$ = 1.89 $\Omega$, and $R_{1'2',4'3'}$ = 0.48 $\Omega$. Above $H$ = 1.3 T, all four resistances have weak negative MR. Below $H$ = 1.3 T, there is an ($x$-$y$) anisotropy between the MR with current directed $\parallel\bf{\hat{x}}$ (local minima) and $\parallel\bf{\hat{y}}$ (maxima), likely due to domain-dependent scattering. Interestingly, the MR  $\parallel\bf{\hat{y}}$ has a large $H$-antisymmetric component with the same $\sim$1 T shoulder as the anomalous Hall resistance (panel c). {\bf b.} Exchanging the applied current and voltage contacts, $ij\leftrightarrow kl$, flips the sign of this antisymmetric MR. {\bf c, d.} To measure the Hall effect, we average two transverse resistance configurations on each face, e.g., $R_{13,42}$ and $R_{42,31}$. The raw resistances directly show the 90$^\circ$-twist anisotropy. On the upper face, $R_{13,42}$ has a positive value at $H$ = 0 T (red/black curves). On the lower face, $R_{4'2',3'1'}$ (this configuration is rotated by 90$^\circ$)  has roughly the same positive offset. The raw curves show sharp hysteric peaks near the coercive field ($\simeq$ 30 Oe) that provide evidence for the low-field removal of domain walls. The field-sweeping directions are indicated by the colored arrows. Panel d shows how after averaging the two transverse configurations together, the intrinsic Hall resistance is exactly the same in either face.
    }
    \label{fig:fig3}
\end{figure*}

\begin{figure*}
    \centering
  \includegraphics[width=\textwidth]{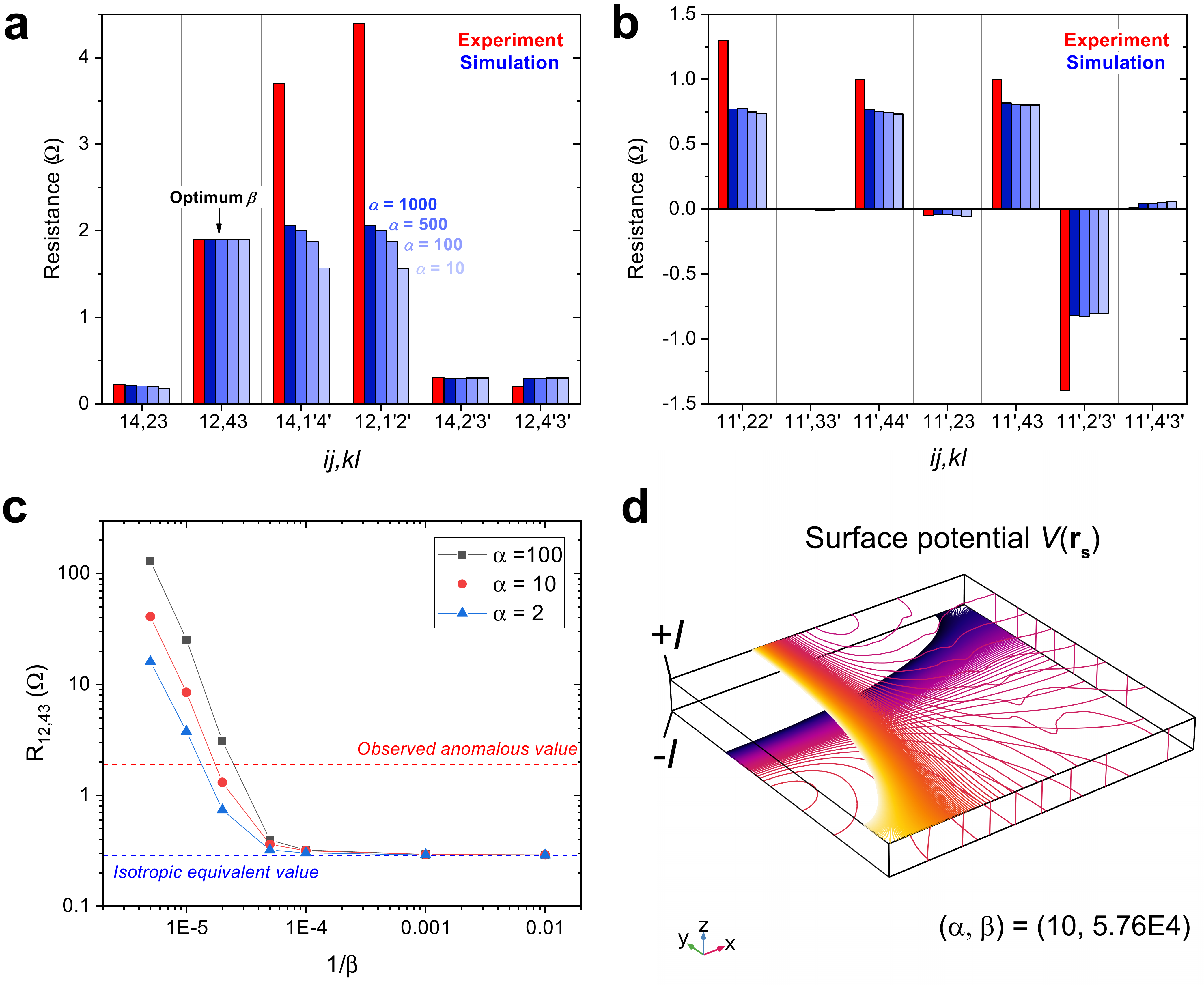}
    \caption{{\bf Simulation results for a model with ultrathin anisotropic surfaces. a, b.}
    Comparison of the experimental (red) and simulated (shades of blue) resistances in the $\delta$ = 10 nm thin-surface model. For a range of $\alpha$ = $\{$10, 100, 500, 1000$\}$, we fine-tuned $\beta$ so that the simulated $R_{12,43}$ (low-conductance-axis) matches the experimental value within 0.5\%. The optimal $\beta$ were found to be 5.76, 4.31, 3.65, and 3.4 $\times$ 10$^4$, respectively.  Panel a (b) presents the results with the current contacts $\vec{ij}$ $\parallel\bf{\hat{x},\hat{y}}$ ($\parallel\bf{\hat{z}}$). Due to the $C_4I$ symmetry, it suffices to simulate only $\frac14$ of the possible four-point resistance configurations. {\bf c.} This plot shows that although the experimentally observed anomalous resistances can be recreated for a range of $\alpha$, no anisotropy arises in this model for $\beta$ $<$ 10$^4$. Without a large suppression of the out-of-plane conductivity in the surface sections, the anomalous resistances revert to the expected isotropic-equivalent values. {\bf d.} This diagram shows the surface potential contours that arise when current is directed out-of-plane, $ij$ = 11'. In order to resolve the non-local voltages, values greater than $\pm$ 1 $\Omega$ from the $C_4I$-symmetric value at the midpoint of $33'$ are deliberately cut off. The spacing between contours is 0.01 $\Omega$. The contours appear to have discrete jumps between the two faces: this is because the thickness of the ultrathin anisotropic surfaces cannot be resolved (1\% of the total slab thickness).
    }
    \label{fig:fig4}
\end{figure*}
\begin{figure*}
    \centering
  \includegraphics[width=\textwidth]{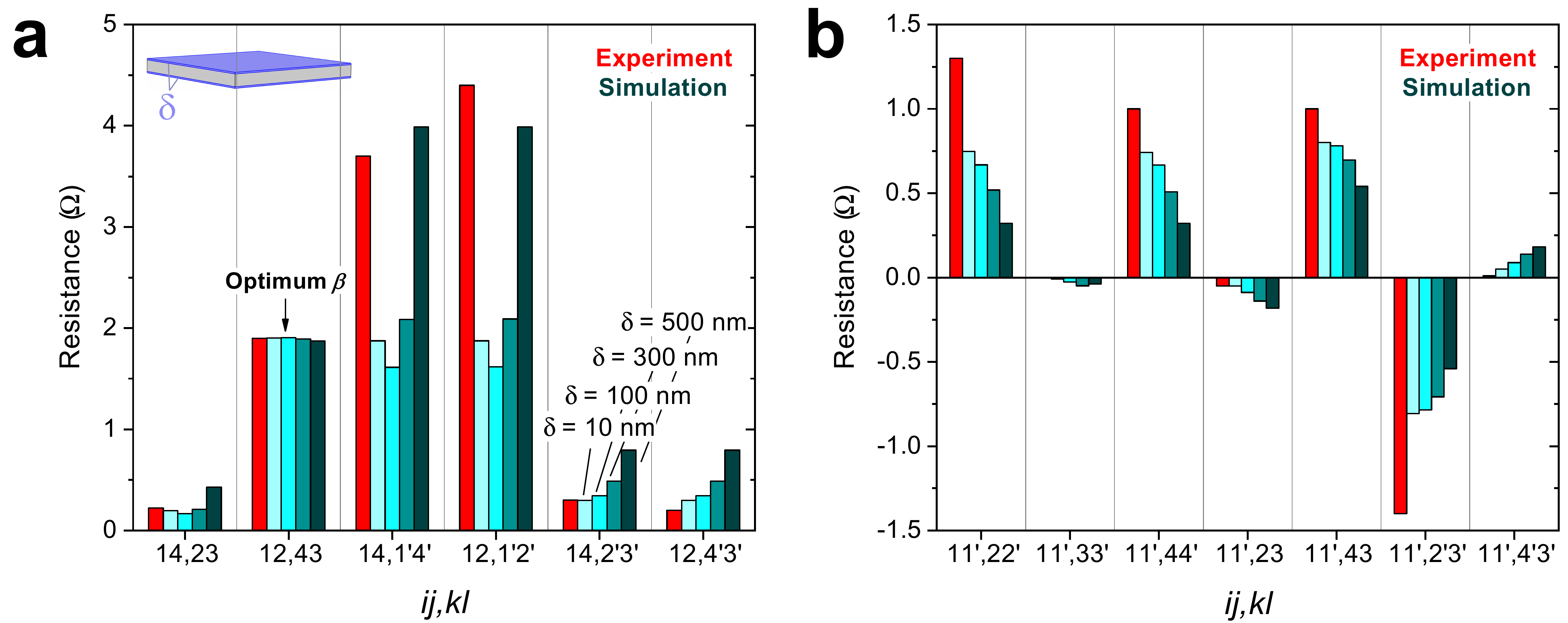}
    \caption{{\bf Thicker-surface simulation results.} We vary the thickness of the anisotropic surface sections $\delta$ and tune $\beta$ to match the experimental value of $R_{12,43}$ = 1.9 $\Omega$ ($\alpha$ is fixed at 100). The red columns represent the experimental resistances and the teal columns show the simulation results. Panel a (b) presents the results with the current contacts $\vec{ij}$ $\parallel\bf{\hat{x},\hat{y}}$ ($\parallel\bf{\hat{z}}$). As $\delta$ is increased, the simulation results largely diverge from the experiment. In particular, the high-conductance-axis configurations ($ij$ = 14) no longer match those of an isotropic slab---a key feature of the experimental anisotropy---and the vertical-current non-local resistance anisotropy becomes less extreme.
    }
    \label{fig:fig5}
\end{figure*}

\end{document}